\documentstyle[12pt]{article}
\makeatletter

\@addtoreset{equation}{section}

\makeatother
\makeatletter
\def\abstract#1{\long\def\@abstract{#1}}%
\def\@abstract{}%
\let\@oldmaketitle=\@maketitle%
\def\@maketitle{%
\@oldmaketitle%
\begin{center}\large\bf Abstract\end{center}%
\begin{quotation}\@abstract\end{quotation}%
\vskip 1.5em}%
\makeatother

\newcommand{\fxi}{\mbox{\boldmath$\xi$}}
\newcommand{\feta}{\mbox{\boldmath$\eta$}}

\newcommand{\ffm}{\mbox{\boldmath$m$}}
\newcommand{\ffl}{\mbox{\boldmath$l$}}
\newcommand{\ffn}{\mbox{\boldmath$n$}}

\begin{document}

\title{EXACTNESS IN THE WKB APPROXIMATION FOR SOME HOMOGENEOUS SPACES }
\author{Kunio FUNAHASHI, Taro KASHIWA, Seiji SAKODA\\
Department of Physics, Kyushu University, Fukuoka 812, Japan\\and\\
Kazuyuki FUJII\\
Department of Mathematics,\\ Yokohama City University, Yokohama 236, Japan}
\abstract{
Analysis of the WKB exactness in some homogeneous spaces is attempted.
$CP^N$ as well as its noncompact counterpart $D_{N,1}$ is studied.
$U(N\!+\!1)$ or $U(N,1)$ based on the Schwinger bosons leads us to
$CP^N$ or $D_{N,1}$ path integral expression for the quantity, ${\rm tr}
e^{-iHT}$, with the aid of coherent states. The WKB approximation terminates
in the leading order and yields the exact result provided that the
Hamiltonian is given by a bilinear form of the creation and the annihilation
operators.
An argument on the WKB exactness to more general cases is also made.
}
\maketitle\thispagestyle{empty}
\newpage

\section{Introduction}\label{sec:jo}

The WKB approximation as the saddle point method in path integral seems
most handy and popular.
However when the exponent (action), under the path integral formula,
consists of quadratic forms, to wit, gives a Gaussian integral, it results
in an exact answer: the harmonic oscillator is the only example so far.
A new possibility, inspired by the theorem of Duistermaat-Heckman\cite{DH,FK},
has recently opened up: quantum mechanical system on non-trivial manifolds,
such as $CP^1$, $CP^N$, and Grassmannian, have been attacked\cite{RRS,MSM}
and seem to possess this property.
The discussions are based on classical as well as geometrical actions
in path integral as a direct consequence of (naive) use of coherent states
to convert operators into $c$-numbers\cite{MS,KNST}: an approximation has
been employed that $\langle g\vert g^\prime\rangle\sim1+\langle g\vert
\delta g\rangle\sim\exp\langle g\vert\delta g\rangle$ where $\vert g
\rangle$ is some (generalized) coherent state and $g^\prime$ is assumed
that $g^\prime\!=\!g+\delta g,\ \delta g\!\ll\!1.$ However this
cannot be legitimate under path integral since $g$ and $g'$ are the
integration variables.
After adopting this approximation, it must be noted that the
{\em resultant action has already been semiclassical.}

With these in mind we discussed $CP^1$, $SU(2)$-spin system, as well as its
noncompact counterpart $SU(1,1)$ in the foregoing paper\cite{FKSF} to confirm
that the expectation does hold indeed. We here extend the survey to $CP^N$ and
its noncompact counterpart $D_{N,1}$ to establish the exactness of the WKB
approximation.

The plan of this paper is as follows: in section 2, a brief introduction to
the generalized coherent states\cite{KS,AP,WZ} based on the
Schwinger bosons is given to set up the trace formula of $CP^N$ system. In
sections 3 and 4, the WKB approximation is explicitly performed to confirm
that
there is no higher order corrections. The subsequent section 5 is devoted to
analyze that the result obtained through the WKB is indeed exact.
The case for a non-compact manifold is picked up in section 6.
The final section 7 is the discussion where the reason of the exactness is
clarified to open the possibility to more general cases.

\section{Coherent States and The Trace Formula}\label{sec:kousei}

We construct the coherent state of $CP^N$ system, in terms of Schwinger
boson formalism~\cite{JS}.
First we consider the system which consists of $N+1$ harmonic oscillators.
The commutation relations are
\begin{equation}
\left[ a_\alpha,a^\dagger_\beta\right]=\delta_{\alpha\beta}\ ,
\left[ a_\alpha,a_\beta\right]=\left[
a^\dagger_\alpha,a^\dagger_\beta\right]=0
\ ,\ \left(\alpha,\beta=1,\cdots N+1\right)\ ,\label{shindoushi}
\end{equation}
and the Fock space is
\begin{eqnarray}
&&\left\{\vert n_1,\cdots,n_{N+1}\rangle\right\}\ ,\ \left( n_\alpha=0,1,2,
\cdots\ {\rm with}\ \alpha=1,\cdots,N+1\right)\ ,\nonumber\\
&&\vert n_1,\cdots,n_{N+1}\rangle\equiv{1\over\sqrt{n_1!\cdots n_{N+1}!}}
\left( a^\dagger_1\right)^{n_1}\cdots\left( a^\dagger_{N+1}\right)^{n_{N+1}}
\vert0\rangle\ ,\nonumber\\
&&a_\alpha\vert0\rangle=0\ .\label{fock}
\end{eqnarray}
By putting
\begin{equation}
E_{\alpha\beta}=a^\dagger_\alpha a_\beta\ ,\ \left(\alpha,\beta=1,\cdots,
N+1\right)\ ,\label{ejitsugen}
\end{equation}
$u(N+1)$ algebra is realized
\begin{eqnarray}
\left[ E_{\alpha\beta},E_{\gamma\delta}\right]=\delta_{\beta\gamma}
E_{\alpha\delta}-\delta_{\delta\alpha}E_{\gamma\beta}\ ,\
\left(\alpha,\beta,\gamma,\delta=1,\cdots,N+1\right)\ .\label{jitsugen}
\end{eqnarray}
Introducing the projection operator
\begin{equation}
P_Q\equiv\int^{2\pi}_0{d\lambda\over2\pi}e^{i\lambda\left({\bf a}^\dagger
{\bf a}-Q\right)}\ ,\label{shaei}
\end{equation}
with
\begin{equation}
{\bf a}\equiv\left( a_1,\cdots,a_{N+1}\right)^T\ ,
\end{equation}
which can be rewritten as
\begin{equation}
P_Q=\sum_{\| n\|=Q}\vert n_1,\cdots,n_{N+1}\rangle\langle n_1,
\cdots,n_{N+1}\vert\ ,
\end{equation}
where $\| n\|=Q$ designates $\sum^{N+1}_{\alpha=1}n_\alpha=Q$.

With the aid of the resolution of unity in terms of the canonical
coherent state~\cite{KS}
\begin{equation}
\int{\left[ d{\bf z}^\dagger
d{\bf z}\right]^{N+1}\!\!\!\!\!\!\!\!\over\pi^{N+1}}
\left\vert{\bf z}\rangle\langle{\bf z}\right\vert={\bf 1}
\left(=\sum^\infty_{n_1=0}\cdots\sum^\infty_{n_{N+1}=0}\vert n_1,
\cdots,n_{N+1}\rangle\langle n_1,\cdots,n_{N+1}\vert\right)\ ,
\label{ztaninobunkai}
\end{equation}
where
\begin{equation}
\left\vert{\bf z}\rangle\right.\equiv e^{-{1\over2}{\bf z}^\dagger{\bf z}}
e^{{\bf a}^\dagger{\bf z}}\left\vert0\rangle\right.\ ,\
{\bf z}\equiv\left( z^1,\cdots,z^N,z^{N+1}\right)^T\in{\bf C}^{N+1}\ ,
\end{equation}
and
\begin{equation}
\left[ d{\bf z}^\dagger d{\bf z}\right]^{N+1}\equiv
\prod^{N+1}_{\alpha=1}d{z^\alpha}^*dz^\alpha\ ,
\end{equation}
$P_Q$ becomes such that
\begin{eqnarray}
P_Q&=&\int{\left[ d{\bf z}^\dagger d{\bf z}\right]^{N+1}\!\!\!\!\!\!\!\!
\over\pi^{N+1}}
\left\vert{\bf z}\rangle\langle{\bf z}\right\vert P_Q\ \nonumber\\
&=&\int\prod^N_{\alpha=1}{d{\xi^\alpha}^*d\xi^\alpha\over\pi}
{\left\vert\zeta\right\vert^{2N}d\zeta^*d\zeta\over\pi\left(1+\fxi^\dagger
\fxi\right)^{N+1}}e^{-\left\vert\zeta\right\vert^2}\nonumber\\
&&\times\sum^\infty_{n=0}\sum_{\| m\|=n}{1\over\sqrt{\ffm!}}
\left({\zeta\over\sqrt{1+\fxi^\dagger\fxi}}\right)^n
\left(\xi^1\right)^{m_1}\cdots\left(\xi^N\right)^{m_N}
\vert m_1,\cdots,m_{N+1}\rangle\nonumber\\
&&\times\sum_{\| l\|=Q}\langle l_1,\cdots,l_{N+1}\vert{1\over\sqrt{\ffl!}}
\left({\zeta^*\over\sqrt{1+\fxi^\dagger\fxi}}\right)^Q
\left({\xi^1}^*\right)^{l_1}\cdots\left({\xi^N}^*\right)^{l_N}\ ,
\label{pzdetenkai}
\end{eqnarray}
where use has been made of a change of variables
\begin{equation}
\pmatrix{z^1\cr\vdots\cr z^N\cr z^{N+1}\cr}
=z^{N+1}\pmatrix{z^1/z^{N+1}\cr\vdots\cr z^N/z^{N+1}\cr1\cr}
\equiv
{\zeta\over\sqrt{1+\fxi^\dagger\fxi}}\pmatrix{\xi^1\cr\vdots\cr\xi^N\cr1\cr}\
,
\ \zeta\equiv\sqrt{{\bf z}^\dagger{\bf z}}\ ,\label{zkaraxi}
\end{equation}
with the assumption that $z^{N+1}\ne0$ to the second line and the abbreviation
\begin{equation}
\ffm!\equiv m_1!\cdots m_{N+1}!\ ,\label{futomteigi}
\end{equation}
and the notation
\begin{equation}
\fxi\equiv\left(\xi^1,\cdots,\xi^N\right)^T\in{\bf C}^N\ ,
\end{equation}
has been adopted.
After the integration with respect to $\zeta$, $P_Q$ can further be cast into
the form  such that
\begin{eqnarray}
P_Q&=&\int\prod^N_{\alpha=1}{d{\xi^\alpha}^*d\xi^\alpha\over\pi}
{\left( N+Q\right)!\over\left(1+\fxi^\dagger\fxi\right)^{N+Q+1}}
\sum_{\| m\|=Q}{1\over\sqrt{\ffm!}}
\left(\xi^1\right)^{m_1}\cdots\left(\xi^N\right)^{m_N}
\vert m_1,\cdots,m_{N+1}\rangle\nonumber\\
&&\times\sum_{\| l\|=Q}\langle l_1,\cdots,l_{N+1}\vert{1\over\sqrt{\ffl!}}
\left({\xi^1}^*\right)^{l_1}\cdots\left({\xi^N}^*\right)^{l_N}\nonumber\\
&=&\int d\mu\left(\fxi,\fxi^\dagger\right)\left\vert\fxi\rangle
\langle\fxi\right\vert\ ,\label{shaeis}
\end{eqnarray}
where
\begin{equation}
\vert\fxi\rangle\equiv{1\over\left(1+\fxi^\dagger\fxi\right)^{Q/2}}
\sum_{\| n\|=Q}\sqrt{Q!\over\ffn!}
\left(\xi^1\right)^{n_1}\cdots\left(\xi^N\right)^{n_N}
\vert n_1,\cdots,n_{N+1}\rangle\ ,\label{joutai}
\end{equation}
and
\begin{equation}
d\mu\left(\fxi,\fxi^\dagger\right)\equiv{\left(
Q+N\right)!\over Q!} {\left[ d\fxi^\dagger
d\fxi\right]^N\over\pi^N
\left(1+\fxi^\dagger\fxi\right)^{N+1}}\ ,\label{measure}
\end{equation}
with
\begin{equation}
\left[ d\fxi^\dagger d\fxi\right]^N\equiv
\prod^N_{\alpha=1}d{\xi^\alpha}^*d\xi^\alpha\ .
\end{equation}
The inner product of $\vert\fxi\rangle$'s is given by
\begin{equation}
\langle\fxi\vert\fxi^\prime\rangle={\left(1+\fxi^\dagger\fxi^\prime\right)^Q
\over\left(1+\fxi^\dagger\fxi\right)^{Q/2}\left(1+{\fxi^\prime}^\dagger
\fxi^\prime\right)^{Q/2}}\ .\label{naiseki}
\end{equation}

So far we confine ourselves in the Fock representation which is now
dictated in terms of $U(N+1)$ representation:
the highest weight vector, defined by
\begin{equation}
E_{N+1,N+1}\vert Q;N+1\rangle\rangle=Q\vert Q;N+1\rangle\rangle\ ,\
E_{N+1,\alpha}\vert Q;N+1\rangle\rangle=0\ ,
\end{equation}
is identified such that
\begin{equation}
\vert Q;N+1\rangle\rangle\equiv\vert\stackrel{1}{0},\cdots,\stackrel{N}{0},
\stackrel{N+1}{Q}\rangle\ .\label{omomi}
\end{equation}
Thus $\vert\fxi\rangle$ can be regarded as the
generalized coherent state
\begin{equation}
\vert\fxi\rangle={1\over\left(1+\fxi^\dagger\fxi\right)^{Q/2}}
e^{\xi^\alpha E_{\alpha,N+1}}\vert Q;N+1\rangle\rangle\ ,\label{cohteigi}
\end{equation}
because the right-hand side is rewritten as
\begin{eqnarray}
{\rm R.H.S.\ of}\ \left(\ref{cohteigi}\right)&=&
{1\over\left(1+\fxi^\dagger\fxi\right)^{Q/2}}
\sum^\infty_{n=1}{1\over n!}\left(\sum^N_{\alpha=1}\xi^\alpha
a^\dagger_\alpha a_{N+1}\right)^n\vert Q;N+1\rangle\rangle\nonumber\\
&=&{1\over\left(1+\fxi^\dagger\fxi\right)^{Q/2}}\sum_{\| n\|=Q}
\left(\xi^1\right)^{n_1}\cdots\left(\xi^N\right)^{n_N}
\vert n_1,\cdots,n_{N+1}\rangle\ ,
\end{eqnarray}
where $E_{\alpha,N+1}(E_{N+1,\alpha})$ is the lowering (raising) operator
of $u(N+1)$ (\ref{ejitsugen}) and use has been made of the explicit form
(\ref{jitsugen}).
It should be noted that the change of variables (\ref{zkaraxi}) corresponds
to picking up the highest weight (\ref{omomi}) (, which can be seen by putting
$\fxi=0$ in (\ref{joutai})).
Another highest weight is also available such that
\begin{equation}
\vert Q;I\rangle\rangle\equiv\vert\stackrel{1}{0},\cdots,
\stackrel{I-1}{0},\stackrel{I}{Q},\stackrel{I+1}{0},\cdots,\stackrel{N+1}{0}
\rangle\ .\label{isaikou}
\end{equation}
(There are $N+1$ highest weight vectors in this sense.)
With this identification, we can assign the projection operator as
the resolution of unity
\begin{equation}
{\bf 1}_Q\equiv P_Q=\int d\mu\left(\fxi,\fxi^\dagger\right)
\vert\fxi\rangle\langle\fxi\vert\ .\label{tani}
\end{equation}

Take a Hamiltonian
\begin{equation}
\hat H\equiv{\bf a}^\dagger H{\bf a}\ ,
\end{equation}
with
\begin{equation}
H\equiv\pmatrix{c_1&&&\cr&c_2&&\cr&&\ddots&\cr&&&c_{N+1}\cr}\ ,
\label{hamiltonian}
\end{equation}
where we have assumed that all $c$'s are different from each other.
The matrix element is calculated to be
\begin{equation}
\langle\fxi\vert\hat H\vert\fxi^\prime\rangle=Q\langle\fxi\vert
\fxi^\prime\rangle{c_{N+1}+\sum^N_{\alpha=1}c_\alpha{\xi^\alpha}^*
{\xi^\prime}^\alpha\over1+\fxi^\dagger\fxi^\prime}\ .\label{meofham}
\end{equation}
In the following we concentrate on the quantity,
\begin{equation}
Z\equiv{\rm tr}\left( e^{-i\hat HT}\right)\equiv\int d\mu\left(\fxi,
\fxi^\dagger\right)\langle\fxi\vert e^{-i\hat HT}\vert\fxi\rangle
\ ,\label{teigitr}
\end{equation}
called the trace formula, which turns out to be
\begin{eqnarray}
Z&=&\lim_{M\to\infty}\int d\mu\left(\fxi,\fxi^\dagger\right)
\langle\fxi\vert\left(1-i\hat H\varepsilon\right)^M\vert\fxi\rangle
\ ,\ \ \left( \varepsilon\equiv{T\over M}\right)\nonumber\\
&=&\lim_{M\to\infty}\int_{\rm PBC}\prod^M_{i=1}d\mu\left(\fxi_i,\fxi^\dagger_i
\right)\prod^M_{k=1}\langle\fxi_k\vert\left(1-i\hat H\varepsilon\right)\vert
\fxi_{k-1}\rangle\nonumber\\
&=&\lim_{M\to\infty}\int_{\rm PBC}\prod^M_{i=1}d\mu\left(\fxi_i,\fxi^\dagger_i
\right)\prod^M_{k=1}\langle\fxi_k\vert\fxi_{k-1}\rangle
\left(1-i\varepsilon{\langle\fxi_k\vert\hat H\vert\fxi_{k-1}\rangle\over
\langle\fxi_k\vert\fxi_{k-1}\rangle}\right)\nonumber\\
&=&\lim_{M\to\infty}\int_{\rm PBC}\prod^M_{i=1}d\mu\left(\fxi_i,\fxi^\dagger_i
\right)\prod^M_{k=1}\langle\fxi_k\vert\fxi_{k-1}\rangle
\exp\left(-i\varepsilon{\langle\fxi_k\vert\hat H\vert\fxi_{k-1}\rangle\over
\langle\fxi_k\vert\fxi_{k-1}\rangle}\right)\ ,\label{risanka}
\end{eqnarray}
where we have used the definition of the exponential function in the first
line, and the resolution of unity (\ref{tani}) in the second line and discarded

$O(\varepsilon^2)$ terms in the final expression and PBC (periodic boundary
condition) designates
\begin{equation}
\fxi_0=\fxi_M\ ;\ ({\rm PBC})\ .\label{shuuki}
\end{equation}
exponentiate the inner product $\prod^M_{k=1}\langle\fxi_k\vert\fxi_{k-1}
\rangle$ to find
\begin{eqnarray}
Z&=&\lim_{M\to\infty}\int_{\rm
PBC}\prod^M_{i=1}d\mu\left(\fxi_i,\fxi^\dagger_i
\right)\exp\left\{\sum^M_{k=1}\left(\log\langle\fxi_k\vert\fxi_{k-1}\rangle
-i\varepsilon{\langle\fxi_k\vert\hat H\vert\fxi_{k-1}\rangle\over
\langle\fxi_k\vert\fxi_{k-1}\rangle}\right)\right\}\nonumber\\
&=&\lim_{M\to\infty}\int_{\rm PBC}\prod^M_{i=1}d\mu\left(\fxi_i,\fxi^\dagger_i
\right)\exp\Bigg[\sum^M_{k=1}\bigg\{\log{\left(1+\fxi^\dagger_k\fxi_{k-1}
\right)^Q\over\left(1+\fxi^\dagger_k\fxi_k\right)^{Q/2}
\left(1+\fxi^\dagger_{k-1}\fxi_{k-1}\right)^{Q/2}}\nonumber\\
&&-i\varepsilon Q{c_{N+1}+\sum^N_{\alpha=1}c_\alpha
{\xi^\alpha_k}^*\xi^\alpha_{k-1}\over1+\fxi^\dagger_k\fxi_{k-1}}
\bigg\}\Bigg]\nonumber\\
&=&e^{-iQc_{N+1}T}\lim_{M\to\infty}\int_{\rm PBC}\prod^M_{i=1}
d\mu\left(\fxi_i,\fxi^\dagger_i\right)
\exp\left[ Q\left\{-\sum^M_{k=1}\log\left(1+\fxi^\dagger_k\fxi_k\right)
\right.\right.\nonumber\\
&&\left.\left.+\sum^M_{k=1}\log\left(1+\fxi^\dagger_k\fxi_{k-1}\right)
-i\varepsilon\sum^M_{k=1}{\sum^N_{\alpha=1}\mu_\alpha{\xi^\alpha_k}^*
\xi^\alpha_{k-1}\over1+\fxi^\dagger_k\fxi_{k-1}}\right\}\right]\ ,
\end{eqnarray}
where (\ref{naiseki}) and (\ref{meofham}) have been applied to give the
second line and $\mu_\alpha\equiv c_\alpha-c_{N+1}$ in the last
expression.
By noting that $\log(1+x)\simeq x$ and discarding $O(\varepsilon^2)$
terms again, the trace formula (\ref{teigitr}) yields to
\begin{eqnarray}
Z_{N+1}&=&e^{-iQc_{N+1}T}\lim_{M\to\infty}\int_{\rm PBC}\prod^M_{i=1}
d\mu\left(\fxi_i,\fxi^\dagger_i\right)\exp\left[ Q\left\{
-\sum^M_{k=1}\log\left(1+\fxi^\dagger_k\fxi_k\right)\right.\right.\nonumber\\
&&\left.\left.+\sum^M_{k=1}\log\left(1+\fxi^\dagger_k\fxi_{k-1}\right)
+\sum^M_{k=1}\log\left(1-i\varepsilon{\sum^N_{\alpha=1}\mu_\alpha
{\xi^\alpha_k}^*\xi^\alpha_{k-1}\over1+\fxi^\dagger_k\fxi_{k-1}}\right)
\right\}\right]\ ,
\end{eqnarray}
which is further rewritten again by use of the rule of discarding
$O(\varepsilon^2)$ terms to
\begin{eqnarray}
Z_{N+1}&=&e^{-iQc_{N+1}T}\lim_{M\to\infty}\int_{\rm PBC}\prod^M_{i=1}d\mu
\left(\fxi_i,\fxi^\dagger_i\right)\nonumber\\
&&\times\exp\left[Q\!\left\{-\sum^M_{k=1}\log\!\left(1+\fxi^\dagger_k\fxi_k
\right)\!+\!\sum^M_{k=1}\log\left(1\!+\!\sum^N_{\alpha=1}
e^{-i\varepsilon\mu_\alpha}
{\xi^\alpha_k}^*\xi^\alpha_{k-1}\right)\right\}\right]\ .\label{traceform}
\end{eqnarray}
Here we have written $Z_{N+1}$ for $Z$ to emphasize the subscript of
$c_{N+1}$.

As was mentioned before, a change of variables (\ref{zkaraxi}) corresponds
to choosing out the highest weight $\vert Q;N+1\rangle\rangle$.
If we make another transformation such that
\begin{equation}
\pmatrix{z^1\cr\vdots\cr z^I\cr\vdots\cr z^{N+1}\cr}
=z^I\pmatrix{z^1/z^I\cr\vdots\cr1\cr\vdots\cr z^{N+1}/z^I\cr}
\equiv{\zeta\over\sqrt{1+\feta^\dagger\feta}}
\pmatrix{\eta^1\cr\vdots\cr1\cr\vdots\cr\eta^{N+1}\cr}\ ,
\end{equation}
with $z^I\ne0$, we obtain
\begin{eqnarray}
\vert\feta\rangle&=&{1\over\left(1+\feta^\dagger\feta\right)^{Q/2}}
\sum_{\| n\|=Q}\sqrt{{Q!\over\ffn!}}\nonumber\\
&&\times\left(\eta^1\right)^{n_1}\cdots\left(\eta^{I-1}\right)^{n_{I-1}}
\left(\eta^{I+1}\right)^{n_{I+1}}\cdots\left(\eta^{N+1}\right)^{n_{N+1}}
\vert n_1,\cdots,n_{N+1}\rangle\ ,
\end{eqnarray}
where
\begin{equation}
\feta\equiv\left(\eta^1,\cdots,\eta^{I-1},\eta^{I+1},\cdots,\eta^{N+1}
\right)^T\ ,
\end{equation}
whose highest weight is (\ref{isaikou}), instead of $\vert
Q;N+1\rangle\rangle$
(\ref{omomi}).
The resolution of unity
\begin{equation}
\int d\mu\left(\feta,\feta^\dagger\right)\vert\feta\rangle\langle\feta\vert=
{\bf 1}_Q\ ,\label{bunkaieta}
\end{equation}
is satisfied.
The trace formula under this coordinate is
\begin{eqnarray}
Z_I&=&e^{-iQc_IT}\lim_{M\to\infty}\int_{\rm PBC}\prod^M_{i=1}
{\left( Q+N\right)!\over Q!}{\left[ d\feta^\dagger_id\feta_i\right]^N\over
\pi^N\left(1+\feta^\dagger_i\feta_i\right)^{N+1}}\nonumber\\
&&\times\exp\Bigg[ Q\bigg\{-\sum^M_{k=1}\log\left(1+\feta^\dagger_k\feta_k
\right)\nonumber\\
&&+\sum^M_{k=1}\log\bigg\{1+e^{-i\varepsilon\left( c_{N+1}-c_I\right)}
{\eta^I_k}^*\eta^I_{k-1}+\sum_{\alpha\ne I}e^{-i\varepsilon\left( c_\alpha
-c_I\right)}{\eta^\alpha_k}^*\eta^\alpha_{k-1}\bigg\}\bigg\}\Bigg]
\ ,\label{etanoz}
\end{eqnarray}
where we have added the subscript $I$ to $Z$.
Comparing $Z_I$ with $Z_{N+1}$,
\begin{eqnarray}
Z_{N+1}&=&e^{-iQc_{N+1}T}\lim_{M\to\infty}\int_{\rm PBC}\prod^M_{i=1}
{\left( Q+N\right)!\over Q!}{\left[ d\fxi^\dagger_id\fxi_i\right]^N\over
\pi^N\left(1+\fxi^\dagger_i\fxi_i\right)^{N+1}}\label{ageta}\\
&&\times\exp\left[Q\left\{-\sum^M_{k=1}\log\left(1+\fxi^\dagger_k\fxi_k\right)
+\sum^M_{k=1}\log\left(1+\sum^N_{\alpha=1}e^{-i\varepsilon\left( c_\alpha
-c_{N+1}\right)}{\xi^\alpha_k}^*\xi^\alpha_{k-1}\right)\right\}\right]\ ,
\nonumber
\end{eqnarray}
we find that $Z_I$ and $Z_{N+1}$ can be interchanged each other by replacing
the subscript $I\leftrightarrow N+1$; which can be understood by the
change of variables
\begin{equation}
\eta^I={1\over\xi^I}\ ,\ \eta^\alpha={\xi^\alpha\over\xi^I} \
\left(\alpha\ne I\right)\ .\label{shinzahyou}
\end{equation}

\section{The WKB Approximation}\label{sec:wkbjikkou}

When $Q$ becomes large in (\ref{etanoz}) or (\ref{ageta}),
the saddle point method is applicable to find
\begin{equation}
\cases{\displaystyle
-{\xi^\alpha_k\over1+\sum^N_{\beta=1}{\xi^\beta_k}^*\xi^\beta_k}
+{e^{-i\varepsilon\mu_\alpha}\xi^\alpha_{k-1}\over1+\sum^N_{\beta=1}
e^{-i\varepsilon\mu_\beta}{\xi^\beta_k}^*\xi^\beta_{k-1}}=0\ ,\cr\cr\cr
\displaystyle
-{{\xi^\alpha_k}^*\over1+\sum^N_{\beta=1}{\xi^\beta_k}^*\xi^\beta_k}
+{e^{-i\varepsilon\mu_\alpha}{\xi^\alpha_{k+1}}^*\over1+\sum^N_{\beta=1}
e^{-i\varepsilon\mu_\beta}{\xi^\beta_{k+1}}^*\xi^\beta_k}=0\ ,\cr
}\label{undohoteishiki}
\end{equation}
which are designated as the equations of motion.
The solutions met with PBC (\ref{shuuki}) are
\begin{equation}
\xi^\alpha_k={\xi^\alpha_k}^*=0\ ,\hspace{6ex}
{\rm\ for\ all}\ \alpha\ {\rm and\ for\ all}\ k\ ,
\end{equation}
or
\begin{equation}
\xi^\alpha_k={\xi^\alpha_k}^*=\infty\ ,\hspace{6ex}
{\rm\ for\ some}\ \alpha\ {\rm and\ for\ all}
\ k\ .\label{shukikai}
\end{equation}
To handle with the latter case, (\ref{shinzahyou}) can be utilized; since
$\xi^I=\infty$ corresponds to $\eta^I=0$.
Thus it is enough to perform a $1/Q$ expansion around $\xi^\alpha_k=0$
in (\ref{ageta}):
\begin{eqnarray}
Z_{N+1}&=&e^{-iQc_{N+1}T}\lim_{M\to\infty}\int_{\rm PBC}\prod^M_{i=1}
{\left[ d\fxi^\dagger_id\fxi_i\right]^N\over\pi^N}\nonumber\\
&&\times\exp\left\{-\left( Q+N+1\right)\sum^M_{k=1}\log\left(1+
\fxi^\dagger_k\fxi_k\right)\right.\nonumber\\
&&\left.+Q\sum^M_{k=1}\log\left(1+\sum^N_{\alpha=1}e^{-i\varepsilon\mu_\alpha}
{\xi^\alpha_k}^*\xi^\alpha_{k-1}\right)\right\}\ ,
\end{eqnarray}
where $(1+\fxi^\dagger\fxi)^{N+1}$, in the measure, has been exponentiated.
Putting
\begin{equation}
\xi^\alpha_k\to\sqrt{\kappa}\xi^\alpha_k\ ,\left(\kappa\equiv{1\over Q}\right)
\ ,\label{xikaraz}
\end{equation}
and performing a formal expansion of the logarithms, we obtain
\begin{eqnarray}
Z_{N+1}\left(\kappa\right)&\equiv&e^{-iQc_{N+1}T}\lim_{M\to\infty}\int_{\rm
PBC}
\prod^M_{i=1}{\left[ d\fxi^\dagger_id\fxi_i\right]^N\over\pi^N}
\exp\left[-\sum^M_{k=1}\sum^N_{\alpha=1}\left({\xi^\alpha_k}^*\xi^\alpha_k
-e^{-i\varepsilon\mu_\alpha}{\xi^\alpha_k}^*\xi^\alpha_{k-1}\right)\right.
\nonumber\\
& &+\sum^\infty_{n=1}\sum^M_{k=1}{\left(-1\right)^n\over n}\kappa^n
\left[\left\{-\sum^N_{\alpha=1}\alpha^n+\left( N+1\right)\left(
\sum^N_{\alpha=1}{\xi^\alpha_k}^*\xi^\alpha_k\right)^n\right\}\right.
\nonumber\\
& &\left.\left.-{n\over n+1}\left\{\left(\sum^N_{\alpha=1}
{\xi^\alpha_k}^*\xi^\alpha_k\right)^{n+1}-\left(\sum^N_{\alpha=1}
e^{-i\varepsilon\mu_\alpha}{\xi^\alpha_k}^*\xi^\alpha_{k-1}\right)^{n+1}
\right\}\right]\right]\ .\label{tenkai}
\end{eqnarray}
As for the leading contribution, set $\kappa=0$ in (\ref{tenkai}) to find
\begin{equation}
Z_{N+1}\left(0\right)=e^{-iQc_{N+1}T}\prod^N_{\alpha=1}
{1\over1-e^{-i\mu_\alpha T}}\ .\label{hitotsu}
\end{equation}
We must sum up all contributions, that is, all solutions of
(\ref{undohoteishiki}) to give
\begin{equation}
Z_{\rm all}\left(0\right)\equiv\sum^{N+1}_{\alpha=1}Z_\alpha=
\sum^{N+1}_{\alpha=1}{e^{-iQc_\alpha T}\over\prod_{\beta\ne\alpha}
\left\{1-e^{-i\left( c_\beta-c_\alpha\right) T}\right\}}\ .\label{wkbtr}
\end{equation}
In the following, we prove that there is no further corrections to
(\ref{wkbtr}):
\begin{equation}
Z_{\rm all}\left(\kappa\right)=Z_{\rm all}\left(0\right)\ .
\end{equation}

\section{Proof of No Corrections}\label{sec:shoumei}

Rewrite (\ref{tenkai}) as
\begin{eqnarray}
Z_{N+1}\left(\kappa\right)&=&e^{-iQc_{N+1}T}\lim_{M\to\infty}\int_{\rm PBC}
\prod^M_{i=1}{\left[ d\fxi^\dagger_id\fxi_i\right]^N\over\pi^N}
\nonumber\\
&&\times\exp\left[\sum^\infty_{n=1}\sum^M_{k=1}{\left(-1\right)^n\over n}
\kappa^n\left[\left\{-\sum^N_{\alpha=1}\alpha^n+\left( N+1\right)\left(
-\partial_{s_k}\right)^n\right\}\right.\right.\nonumber\\
&&\left.\left.-{n\over n+1}\left\{\left(-\partial_{s_k}\right)^{n+1}
-\left(\partial_{t_k}\right)^{n+1}
\right\}\right]\right]\nonumber\\
&&\times\exp\left\{-\sum^M_{k=1}\left( s_k{\xi^\alpha_k}^*\xi^\alpha_k
-t_ke^{-i\varepsilon\mu_\alpha}{\xi^\alpha_k}^*\xi^\alpha_{k-1}\right)\right\}
\nonumber\\
&=&\left. e^{-iQc_{N+1}T}\lim_{M\to\infty}\prod^M_{i=1}
F\left(-\partial_{s_i},\partial_{t_i}\right)
G\left({\bf s},{\bf t}\right)\right\vert_{\left\{ s\right\}
=\left\{ t\right\}=1}\ .\label{shinkigou}
\end{eqnarray}
Here and hereafter we use the following notations
\begin{eqnarray}
F\left( x,y\right)&\equiv&\exp\!\left[\sum^\infty_{n=1}
{\left(-1\right)^n\over n}\kappa^n\!\left\{\!-\!\sum^N_{\alpha=1}\alpha^n
\!+\!\left( N+1\right) x^n\!-\!{n\over n+1}\left( x^{n+1}-y^{n+1}\right)
\right\}\right]\ ,\\
G\left( s,t\right)&\equiv&\int^\infty_0\prod^N_{\alpha=1}d\tau_\alpha
\exp\left\{-\sum^N_{\alpha=1}\left( s-e^{-i\mu_\alpha T}t\right)
\tau_\alpha\right\}\ ,\\
\left\{ s\right\}=1&\Leftrightarrow&\left\{s_1=1,\cdots,s_M=1\right\}\ ,\\
{\bf s}&\equiv&s_1s_2\cdots s_M\ ,\\
\hat{\bf s}_i&\equiv&s_1\cdots s_{i-1}s_{i+1}\cdots s_M=\prod_{j\ne i}s_j
\ ,\\
\hat{\left\{ s\right\}}_i&\equiv&\left\{ s\right\}\setminus s_i\ .
\label{kigourui}
\end{eqnarray}
The next formula plays a central role.

\begin{flushleft}{\bf Formula:}\end{flushleft}
\begin{equation}
\left.F\left(-\partial_{s_i},\partial_{t_i}\right)
G\left({\bf s},{\bf t}\right)\right\vert_{s_i=t_i=1}
=G\left(\hat{\bf s}_i,\hat{\bf t}_i\right)\ .\label{koushiki}
\end{equation}
This formula states that an application of
$\left.F\left(-\partial_{s_i},\partial_{t_i}\right)
\right\vert_{s_i=t_i=1}$ to $G\left({\bf s},{\bf t}\right)$ erases $s_i$ and
$t_i$ from $G\left({\bf s},{\bf t}\right)$.
Thus a repeated use of that leads to
\begin{eqnarray}
Z_{N+1}\left(\kappa\right)&=&\left.e^{-iQc_{N+1}T}\lim_{M\to\infty}\prod_{j\ne
i}
F\left(-\partial_{s_j},\partial_{t_j}\right)
G\left(\hat{\bf s}_i,\hat{\bf t}_i\right)\right\vert_{\hat{\left\{
s\right\}}_i
=\hat{\left\{ t\right\}}_i=1}\nonumber\\
&&\vdots\nonumber\\
&=&\left.e^{-iQc_{N+1}T}F\left(-\partial_s,\partial_t\right)
G\left( s,t\right)\right\vert_{s=t=1}\nonumber\\
&=&e^{-iQc_{N+1}T}G\left(1,1\right)\nonumber\\
&=&e^{-iQc_{N+1}T}\int^\infty_0
\prod^N_{\alpha=1}d\tau\exp\left\{-\sum^N_{\alpha=1}\left(1-e^{-i\mu_\alpha T}
\right)\right\}\nonumber\\
&=&{e^{-iQc_{N+1}T}\over\prod^N_{\alpha=1}
\left(1-e^{-i\mu_\alpha T}\right)}\ .\label{kekkaichi}
\end{eqnarray}

In view of (\ref{kekkaichi}) and (\ref{hitotsu}), we can conclude that
{\em there are no higher order corrections of $\kappa$}.

Now we prove the formula (\ref{koushiki}).
Performing the derivatives in the left-hand side of (\ref{koushiki}),
and putting $s_i=t_i=1$ we find
\begin{eqnarray}
& &\left.F\left(-\partial_{s_i},\partial_{t_i}\right)
G\left({\bf s},{\bf t}\right)\right\vert_{s_i=t_i=1}\nonumber\\
&=&\exp\Bigg[\sum^\infty_{n=1}{\left(-1\right)^n\over n}\kappa^n
\bigg\{-\sum^N_{\alpha=1}\alpha^n+\left( N+1\right)\left(\hat{\bf s}_i
\sum^N_{\alpha=1}\tau_\alpha\right)^n\nonumber\\
&&-{n\over n+1}\left(\left(\hat{\bf s}_i\sum^N_{\alpha=1}
\tau_\alpha\right)^{n+1} -\left( e^{-i\mu_\alpha
T}\hat{\bf t}_i\sum^N_{\alpha=1}\tau_\alpha\right)^{n+1}\right)
\bigg\}\Bigg]\nonumber\\
&&\times\int^\infty_0\prod^N_{\alpha=1}d\tau_\alpha\exp\left\{
-\sum^N_{\alpha=1}\left(\hat{\bf s}_i-e^{-i\mu_\alpha T}\hat{\bf t}_i\right)
\tau_\alpha\right\}\ .\label{bfsayou}
\end{eqnarray}
The right-hand side can be rewritten in the sense of the asymptotic expansion
to give
\begin{eqnarray}
\left(\ref{bfsayou}\right)&\approx&\int^\infty_0\prod^N_{\alpha=1}d\tau_\alpha
\exp\left\{\sum^N_{\alpha=1}\log\left(1+\kappa\alpha\right)-\left( N+1\right)
\log\left(1+\kappa\hat{\bf s}_i\sum^N_{\alpha=1}\tau_\alpha\right)\right.
\nonumber\\
&&\left.-{1\over\kappa}\log\left(1+\kappa\hat{\bf s}_i\sum^N_{\alpha=1}
\tau_\alpha\right)+{1\over\kappa}\log\left(1+\kappa\hat{\bf t}_i
\sum^N_{\alpha=1} e^{-i\mu_\alpha T}\tau_\alpha\right)\right\}\nonumber\\
&=&\int^\infty_0\prod^N_{\alpha=1}d\tau_\alpha\left\{\prod^N_{\alpha=1}
\left(1+\kappa\alpha\right)\right\}{\left(1+\kappa\hat{\bf t}_i
\sum^N_{\alpha=1}e^{-i\mu_\alpha T}
\tau_\alpha\right)^{1/\kappa}\over\left(1+\kappa\hat{\bf s}_i
\sum^N_{\alpha=1}\tau_\alpha\right)^{1/\kappa+N+1}}\
\ .\label{fsayou}
\end{eqnarray}
Adopt the binomial expansion with respect to the numerator as well as
the denominator to find
\begin{eqnarray}
\left(\ref{fsayou}\right)&=&\int^\infty_0\prod^N_{\alpha=1}
d\tau_\alpha\sum^\infty_{n=0}\sum^\infty_{m=0}
\left(\hat{\bf s}_i\sum^N_{\alpha=1}\tau_\alpha\right)^m\left(\hat{\bf t}_i
\sum^N_{\alpha=1}e^{-i\mu_\alpha T}\tau_\alpha\right)^n\nonumber\\
&&\times\kappa^{n+m+N}{\left(-1\right)^m\over n!m!}\prod^{N+m}_{l=-n+1}
\left({1\over\kappa}+l\right)\ .\label{nikou}
\end{eqnarray}
In order to make the $\kappa$-dependence clear, introduce the contour integral

such that
\begin{eqnarray}
\left(\ref{nikou}\right)&=&\sum^\infty_{k=0}{\kappa^k\over2\pi i}
\oint_{\left\vert w\right\vert=1}
{dw\over w^{k+1}}\int^\infty_0\prod^N_{\alpha=1}d\tau_\alpha
\sum^\infty_{n=0}\sum^\infty_{m=0}{\left(-1\right)^m\over n!m!}
\prod^{N+m}_{l=-n+1}\left({1\over w}+l\right)\nonumber\\
& &\times\left(\hat{\bf s}_i\sum^N_{\alpha=1}\tau_\alpha\right)^m
\left(\hat{\bf t}_i\sum^N_{\alpha=1}e^{-i\mu_\alpha T}\tau_\alpha\right)^n
w^{n+m+N}\nonumber\\
&=&\sum^\infty_{k=0}\kappa^k
\int^\infty_0\prod^N_{\alpha=1}d\tau_\alpha\sum^\infty_{n=0}
\sum^\infty_{m=0}{\left(-1\right)^m\over n!m!}\nonumber\\
&&\times\left(\hat{\bf s}_i\sum^N_{\alpha=1}\tau_\alpha\right)^m
\left(\hat{\bf t}_i\sum^N_{\alpha=1}e^{-i\mu_\alpha T}\tau_\alpha\right)^n
{1\over2\pi i}\oint_{\vert w\vert=1}{dw\over w^{k+1}}\prod^{N+m}_{l=-n+1}
\left( lw+1\right)\ .\label{arawa}
\end{eqnarray}
By taking into account that
\begin{eqnarray}
\prod^{N+m}_{l=-n+1}\left( lw+1\right)
&=&\left\{\left( N+m\right) w+1\right\}\left\{\left( N+m-1\right) w+1\right\}
\nonumber\\
&&\times\cdots\times\left\{\left(-n+2\right) w+1\right\}
\left\{\left(-n+1\right) w+1\right\}\nonumber\\
&=&1+\left(\sum^{N+m}_{\beta=-n+1}\beta\right) w+
\left(\sum^{N+m}_{\beta_2=-n+1}\beta_2\sum^{\beta_2-1}_{\beta_1=-n+1}\beta_1
\right) w^2+\cdots\nonumber\\
&&+\left(\sum^{N+m}_{\beta_k=-n+1}\beta_k\sum^{\beta_k-1}_{\beta_{k-1}=-n+1}
\beta_{k-1}\cdots\sum^{\beta_2}_{\beta_1=-n+1}\beta_1\right) w^k+
\cdots\nonumber\\
&&+\left(\sum^{N+m}_{\beta_{N+m+n-1}=-n+1}\beta_{N+m+n-1}\cdots
\sum^{\beta_2}_{\beta_1=-n+1}\beta_1\right) w^{N+m+n-1}\nonumber\\
&&+\left( N+m\right)!\delta_{n0}w^{N+m+n}\ ,
\end{eqnarray}
$w$-integral in (\ref{arawa}) becomes
\begin{eqnarray}
&&{1\over2\pi i}\oint_{\left\vert w\right\vert=1}{dw\over w^{k+1}}
\prod^{N+m}_{l=-n+1}\left( lw+1\right)\nonumber\\
&=&\sum^{N+m}_{\beta_k=-n+1}\beta_k\sum^{\beta_k-1}_{\beta_{k-1}=-n+1}
\beta_{k-1}\cdots\sum^{\beta_2-1}_{\beta_1=-n+1}\beta_1\nonumber\\
&=&\sum^{2k-1}_{l=1}A^{\left( k\right)}_l\!\!\left( n\right)
\sum^{N+m}_{\beta=-n+1}\beta^l\ ,\label{wsekibunwa}
\end{eqnarray}
where we have performed the summation with respect to
$\beta_i\ (i=1\sim k-1)$
leaving coefficients $A^{\left( k\right)}_l\!\!\left( n\right)$;
which emerge in such a way, for example, in $k=1$ case:
\begin{equation}
{\rm\left( The\ 2nd\ line\ of\ \left(\ref{wsekibunwa}\right)\right)}
=\sum^{N+m}_{\beta_1=-n+1}\beta_1\ ,
\end{equation}
therefore $A^{\left(1\right)}_1\!\!\left( n\right)=1$, and in $k=2$ case:
\begin{eqnarray}
{\rm\left( The\ 2nd\ line\ of\ \left(\ref{wsekibunwa}\right)\right)}
&=&\sum^{N+m}_{\beta_2=-n+1}\beta_2\sum^{\beta_2-1}_{\beta_1=-n+1}\beta_1
\nonumber\\
&=&\sum^{N+m}_{\beta_2=-n+1}\left\{{1\over2}\beta^3_2
-{1\over2}\beta_2^2-{1\over2}n\left( n-1\right)\beta_2\right\}\ .
\end{eqnarray}
Thus
\begin{equation}
A^{\left(2\right)}_3\!\!\left( n\right)={1\over2}\ ,\
A^{\left(2\right)}_2\!\!\left( n\right)=-{1\over2}\ ,\
A^{\left(2\right)}_1\!\!\left( n\right)=-{1\over2}n\left( n-1\right)\ ,
\end{equation}
but fortunately these explicit forms for general $k$ are not necessary for
our purpose.
With the aid of the relation,
\begin{eqnarray}
\sum^{N+m}_{\alpha=-n+1}\alpha^k&=&\left.\left({d\over dt}\right)^k
\sum^{N+m}_{\alpha=-n+1}e^{\alpha t}\right\vert_{t=0}\nonumber\\
&=&\left({d\over dt}\right)^k\left.{e^{\left( N+m+1\right) t}-e^{-\left(
n-1\right) t}\over e^t-1}\right\vert_{t=0}\nonumber\\
&=&{k!\over2\pi i}\oint{d\eta\over\eta^{k+1}}{e^{\left( N+m+1\right)\eta}
-e^{-\left( n-1\right)\eta}\over e^\eta-1}\label{wakakinaoshi}\ ,
\end{eqnarray}
we can perform the summation with respect to $m$ in (\ref{arawa}) as follows:
\begin{eqnarray}
&&\left.F\left(-\partial_{s_i},\partial_{t_i}\right) G\left({\bf s},
{\bf t}\right)\right\vert_{s_i=t_i=1}\nonumber\\
&=&{1\over\prod^N_{\alpha=1}\left(\hat{\bf s}_i
-e^{-i\mu_\alpha T}\hat{\bf t}_i\right)}\nonumber\\
&&+\sum^N_{k=1}\kappa^k\prod^N_{\alpha=1}\int^\infty_0d\tau_\alpha
\sum^\infty_{n=0}\sum^\infty_{m=0}{\left(-1\right)^m\over n!m!}
\left(\hat{\bf s}_i\sum^N_{\alpha=1}\tau_\alpha\right)^m\left(\hat{\bf t}_i
\sum^N_{\alpha=1}e^{-i\mu_\alpha T}\tau_\alpha\right)^n\nonumber\\
&&\times\sum^{2k-1}_{l=1} A^{\left( k\right)}_l\left( n\right)
{k!\over2\pi i}\oint{d\eta\over\eta^{k+1}}{e^{\left( N+m+1\right)\eta}
-e^{-\left( n-1\right)\eta}\over e^\eta-1}\nonumber\\
&=&{1\over\prod^N_{\alpha=1}\left(\hat{\bf s}_i
-e^{-i\mu_\alpha T}\hat{\bf t}_i\right)}\nonumber\\
&&+\sum^N_{k=1}\kappa^k\prod^N_{\alpha=1}\int^\infty_0d\tau_\alpha
\sum^\infty_{n=0}{1\over n!}\left(\hat{\bf t}_i\sum^N_{\alpha=1}
e^{-i\mu_\alpha T}\tau_\alpha\right)^n\prod^{2k-1}_{l=1}
A^{\left( k\right)}_l\left( n\right)\nonumber\\
&&\times{k!\over2\pi i}\oint{d\eta\over\eta^{k+1}}{1\over e^{\eta}-1}\left\{
e^{\left( N+1\right)\eta}e^{\hat{\bf s}_ie^\eta\sum^N_{\alpha=1}\tau_\alpha}
-e^{-\left( n-1\right)\eta}e^{-\hat{\bf s}_i\sum^N_{\alpha=1}\tau_\alpha}
\right\}\ ,\label{bunri}
\end{eqnarray}
where we have separated the $0$-th order term from others.
If we notice the relation
\begin{equation}
\int^\infty_0\prod^N_{\alpha=1}d\tau_\alpha{1\over n!}\left( t
\sum^N_{\alpha=1}e^{-i\mu_\alpha T}\tau_\alpha\right)^ne^{-s\sum^N_{\alpha=1}
\tau_\alpha}={t^n\over s^{n+N}}\sum_{\| m_\alpha\|=n}
\prod^N_{\alpha=1}e^{-im_\alpha\mu_\alpha T}\ ,\label{tausekibun}
\end{equation}
obtained from the definition of the Gamma function,
\begin{equation}
\int^\infty_0d\tau\tau^{M-1}e^{-\tau}=\Gamma\left( M\right)\ ,
\end{equation}
(\ref{bunri}) becomes
\begin{eqnarray}
&&\left.F\left(-\partial_{s_i},\partial_{t_i}\right) G\left({\bf s},
{\bf t}\right)\right\vert_{s_i=t_i=1}\nonumber\\
&=&{1\over\prod^N_{\alpha=1}
\left(\hat{\bf s}_i-e^{-i\mu_\alpha T}\hat{\bf t}_i\right)}\nonumber\\
&&+\sum^\infty_{k=1}\kappa^k\sum^\infty_{n=0}\sum^{2k-1}_{l=1}
A^{\left( n\right)}_l\left( n\right){k!\over2\pi i}\oint
{d\eta\over\eta^{k+1}}{1\over e^\eta-1}\nonumber\\
&&\times\hat{\bf t}^n_i\hat{\bf s}^{-\left( n+N\right)}_i
\sum_{\| m_\alpha\|=n}\prod^N_{\alpha=1}e^{-im_\alpha\mu_\alpha T}
\left\{ e^{\left( N+1\right)\eta}
e^{-\left( N+n\right)\eta}-e^{-\left( n-1\right)\eta}\right\}\ ,
\end{eqnarray}
whose second term obviously goes to zero leaving the first term
which is expressed as
\begin{eqnarray}
{1\over\prod^N_{\alpha=1}\left(\hat{\bf s}_i-e^{-i\mu_\alpha T}
\hat{\bf t}_i\right)}
&=&\int^\infty_0\prod^N_{\alpha=1}d\tau_\alpha\exp\left\{-\sum^N_{\alpha=1}
\left(\hat{\bf s}_i-e^{-i\mu_\alpha T}\hat{\bf t}_i\right)\tau_\alpha\right\}
\nonumber\\
&=&G\left(\hat{\bf s}_i,\hat{\bf t}_i\right)\ .\label{saishuu}
\end{eqnarray}
Therefore the formula has been proved.

\section{Exact Calculation}\label{sec:genmitsu}

In the previous sections we see that only the leading order term is surviving.
Thus the next step is to check whether (\ref{wkbtr}) is the correct answer or
not.
To this end, let us make an exact calculation.

In order to absorb the phase factor $e^{-i\varepsilon\mu_\alpha}$
in(\ref{traceform}), a change of variables
\begin{equation}
\xi^\alpha_k\to e^{-ik\varepsilon\mu_\alpha}\xi^\alpha_k\ ,\label{hensuhenkan}
\end{equation}
is made to find that the remnant attributes to the boundary term;
$\xi^\alpha_0=\xi^\alpha_Me^{-i\mu_\alpha T}$.
The trace formula (\ref{traceform}) thus becomes
\begin{eqnarray}
Z&=&e^{-iQc_{N+1}T}\lim_{M\to\infty}\int\prod^M_{i=1}d\mu
\left(\fxi_i,\fxi^\dagger_i\right)\nonumber\\
&&\times\exp\left[-Q\sum^M_{k=1}\log\left(1+\fxi^\dagger_k\fxi_k\right)\right.
+\left.\left. Q\sum^M_{k=1}\log\left(1+\fxi^\dagger_k\fxi_{k-1}\right)
\right]\right\vert_{\xi^\alpha_0=\xi^\alpha_Me^{-i\mu_\alpha T}}
\label{triichi}\\
&=&\left.e^{-iQc_{N+1}}\!\lim_{M\to\infty}\int\!\prod^M_{i=1}
d\mu\!\left(\fxi_i,\fxi^\dagger_i\right)\!
\prod^M_{k=1}{\left(1+\fxi^\dagger_k\fxi_{k-1}\right)^Q\over
\left(1+\fxi^\dagger_k\fxi_k\right)^{Q/2}
\left(1+\fxi^\dagger_{k-1}\fxi_{k-1}\right)^{Q/2}}
\right\vert_{\xi^\alpha_0=\xi^\alpha_Me^{-i\mu_\alpha T}}
\ .\nonumber
\end{eqnarray}
In view of the inner product (\ref{naiseki}), (\ref{triichi}) becomes
\begin{equation}
Z=\left.e^{-iQc_{N+1}T}\lim_{M\to\infty}\int\prod^M_{i=1}d\mu
\left(\fxi_i,\fxi^\dagger_i\right)\prod^M_{k=1}
\langle\fxi_k\vert\fxi_{k-1}\rangle\right\vert_{\xi^\alpha_0=\xi^\alpha_M
e^{-i\mu_\alpha T}}\ ,\label{naisekiseki}
\end{equation}
yielding to
\begin{eqnarray}
Z&=&e^{-iQc_{N+1}T}\int d\mu\left(\fxi,\fxi^\dagger\right)
\langle\fxi\left\vert U\fxi\rangle\right.\nonumber\\
&=&e^{-iQc_{N+1}T}{\left( Q+N\right)!\over Q!\pi^N}\int
{\left[ d\fxi^\dagger d\fxi\right]^N\over\left(1+\fxi^\dagger\fxi
\right)^{N+Q+1}}\left(1+\fxi^\dagger U\fxi\right)^Q\ ,\label{kantan}
\end{eqnarray}
where use have been made of the resolution of unity (\ref{tani}) and
\begin{equation}
U\equiv\pmatrix{e^{-i\mu_1T}&&\cr&e^{-i\mu_2T}&&\cr&&\ddots&\cr&&&e^{-i\mu_NT}
&\cr}\ .\label{mgyouretsu}
\end{equation}
Introducing the ``polar coordinate''
\begin{equation}
\xi^\alpha=\sqrt{u_\alpha}e^{i\theta_\alpha}\ ,\label{kyokuzahyou}
\end{equation}
and integrating out the angular parts, we find
\begin{eqnarray}
Z&=&e^{-iQc_{N+1}T}\left( Q+N\right)!\sum_{\| l\|=Q}
{1\over\ffl!}\prod^N_{\alpha=1}e^{-i\mu_\alpha l_\alpha}\nonumber\\
& &\times\int\prod^N_{\alpha=1}du_\alpha{\left( u_1\right)^{l_1}\cdots
\left( u_N\right)^{l_N}\over\left(1+u_1+\cdots+u_N\right)^{N+Q+1}}\
,\label{doukei}
\end{eqnarray}
where $l!$ is given by (\ref{futomteigi}) with
$l_{N+1}=Q-\sum^N_{\alpha=1}l_\alpha$.
Paying attention to
\begin{eqnarray}
& &\int\prod^N_{\alpha=1}du_\alpha{\left( u_1\right)^{l_1}\cdots
\left( u_N\right)^{l_N}\over\left(1+u_1+\cdots+u_N\right)^{N+Q+1}}\nonumber\\
&=&B\left( l_{N+1},N+Q+1-\left( l_N+1\right)\right)\nonumber\\
&&\times\int du_1\cdots du_{N-1}{\left(u_1\right)^{l_1}\cdots
\left( u_{N-1}\right)^{l_{N-1}}\over
\left(1+u_1+\cdots+u_{N-1}\right)^{N+Q+1-l_{N+1}}}\nonumber\\
&=&{\ffl!\over\left( N+Q\right)!}\ ,\label{usekibunkekka}
\end{eqnarray}
with $B(p,q)$ being the beta-function,
we see that the trace (\ref{doukei}) reads
\begin{equation}
Z=\sum_{\| l\|=Q}e^{-iT\sum^{N+1}_{\alpha=1}c_\alpha l_\alpha}
\ .\label{trikekkaichi}
\end{equation}
This is the result, however, to compare with that of the WKB approximation of
(\ref{wkbtr}), there needs a further modification of (\ref{trikekkaichi}):
\begin{equation}
Z=\lim_{\delta\to0}\int^{2\pi}_0{d\lambda\over2\pi}e^{-i\lambda Q}
\sum^\infty_{l_1=0}\cdots\sum^\infty_{l_{N+1}=0}\exp{\left\{{-iT
\sum^{N+1}_{\alpha=1}c_\alpha l_\alpha +i\left(\lambda+i\delta\right)
\sum^{N+1}_{\alpha=1}l_\alpha}\right\}}\ ,\label{deltaireru}
\end{equation}
where the regularization parameter $\delta$ has been
introduced to control the  infinite series of $l_\alpha$'s.
Therefore after taking $l_\alpha$'s sum we find
\begin{eqnarray}
Z&=&\lim_{\delta\to0}\int^{2\pi}_0{d\lambda\over2\pi}
e^{-i\lambda Q}\prod^{N+1}_{\alpha=1}
{1\over1-e^{-i\left( c_\alpha T-\lambda-i\delta\right)}}\nonumber\\
&=&\lim_{\delta\to0}\oint_{\vert z\vert=1}{dz\over2\pi i}
z^{Q+N}\prod^{N+1}_{\alpha=1} {1\over z-e^{-ic_\alpha T-\delta}}\nonumber\\
&=&\lim_{\delta\to0}\sum^{N+1}_{\alpha=1}
\mathop{\rm Res}_{z=e^{-ic_\alpha T-\delta}}z^{Q+N}
\prod_{\beta\ne\alpha}{1\over z-e^{-ic_\beta T-\delta}}\nonumber\\
&=&\sum^{N+1}_{\alpha=1}{e^{-iQc_\alpha T}\over\prod_{\beta\ne\alpha}
\left\{1-e^{-i\left( c_\beta-c_\alpha\right) T}\right\}}
\ ,\label{genmitsukekka}
\end{eqnarray}
where the $\lambda$-integral has been transformed into the contour integral
in the second line by putting $z=e^{-i\lambda}$.
The result thus coincides with that of the WKB approximation (\ref{wkbtr}),
convincing us that {\em the WKB approximation is exact.}

\section{Noncompact Case}\label{sec:noncompact}

The noncompact cases can be handled with a similar manner as was the case
of $su(1,1)$ discussed in the previous paper~\cite{FKSF}.
$u(N,1)$ algebra is given by
\begin{eqnarray}
&&\left[ E_{\alpha\beta},E_{\gamma\delta}\right]=\eta_{\beta\gamma}
E_{\alpha\delta}-\eta_{\delta\alpha}E_{\gamma\beta}\
,\left(\alpha,\beta,\gamma,\delta=1,\cdots,N+1\right)\ ,
\nonumber\\
&&\eta_{\alpha\beta}={\rm diag}\left(1,\cdots,1,-1\right)\ ,\label{ndaisu}
\end{eqnarray}
with a subsidiary condition
\begin{equation}
-\sum^N_{\alpha=1}E_{\alpha\alpha}+E_{N+1,N+1}=K\ ,\left(
K=N,N+1,\cdots\right)
\ .\label{nkousoku}
\end{equation}

As before $u(N,1)$ algebra is realized in terms of the Schwinger boson
(\ref{shindoushi}), (\ref{fock}):
$E$'s in (\ref{ndaisu}) is
\begin{eqnarray}
&E_{\alpha\beta}=a^\dagger_\alpha a_\beta\ ,
&E_{N+1,N+1}=a^\dagger_{N+1}a_{N+1}+1\ ,\nonumber\\
&E_{N+1,\alpha}=a_{N+1}a_\alpha\ ,
&E_{\alpha,N+1}=a^\dagger_\alpha a^\dagger_{N+1}\ .
\end{eqnarray}
Introducing the projection operator;
\begin{equation}
P_K\equiv\int^{2\pi}_0{d\lambda\over2\pi}
e^{i\lambda\left({\bf a}^\dagger\eta{\bf a}-K+1\right)}\ ,\label{nshaei}
\end{equation}
which can be expressed by
\begin{equation}
P_K=\sum^\infty_{\hat{\| n\|}=0}| n_1,\cdots,n_N,K-1+\hat{\| n\|}
\rangle\langle n_1,\cdots,n_N,K-1+\hat{\| n\|}|\ ,
\end{equation}
where $\hat{\| n\|}\equiv\sum^N_{\alpha=1}n_\alpha$.

Following the same procedure from (\ref{pzdetenkai})$\sim$(\ref{naiseki}),
we have
\begin{equation}
{\bf 1}_K=\int d\mu\left(\fxi,\fxi^\dagger\right)\vert\fxi\rangle
\langle\fxi\vert\ ,\label{ntani}
\end{equation}
where
\begin{eqnarray}
D_{N,1}&\equiv&\left\{\fxi\in{\bf C}^N\vert1-\fxi^\dagger\fxi>0\right\}\ ,
\nonumber\\
\vert\fxi\rangle&\equiv&\left(1-\fxi^\dagger\fxi\right)^{K/2}\sum^\infty_{n=0}
\sum_{\|\hat m\|=n}\sqrt{\left( n+K-1\right)!\over\left( K-1\right)!
{\hat{\ffm}}!}\left(\xi^1\right)^{m_1}\cdots\left(\xi^N\right)^{m_N}
\nonumber\\
&&\times\vert m_1,\cdots,m_N,n+K-1\rangle\ ,\nonumber\\
\fxi&\equiv&\left(\xi^1,\cdots,\xi^N\right)^T\in  D_{N,1}\ ,
\end{eqnarray}
and
\begin{equation}
d\mu\left(\fxi,\fxi^\dagger\right)\equiv{\left( K-1\right)!
\over\left( K-1-N\right)!}{\left[ d\fxi^\dagger d\fxi\right]^N\over
\pi^N\left(1-\fxi^\dagger\fxi\right)^{N+1}}\ ,
\end{equation}
is the invariant measure on $D_{N,1}$.
To derive (\ref{ntani}) we have adopted a change of variables
\begin{equation}
\pmatrix{z^1\cr\vdots\cr z^N\cr z^{N+1}\cr}=\zeta
\pmatrix{\xi^1\cr\vdots\cr\xi^N\cr\sqrt{1-\fxi^\dagger\fxi}\cr}\ .
\end{equation}
The inner product of the states is
\begin{equation}
\langle\fxi\vert\fxi^\prime\rangle={\left(1-\fxi^\dagger\fxi\right)^{K/2}
\left(1-{\fxi^\prime}^\dagger\fxi^\prime\right)^{K/2}\over
\left(1-\fxi^\dagger\fxi^\prime\right)^K}\ .
\end{equation}

The Hamiltonian in this case is
\begin{equation}
H=\sum^{N+1}_{\alpha=1}c_\alpha E_{\alpha\alpha}\ ,
\end{equation}
whose matrix element is
\begin{equation}
\langle\fxi\vert H\vert\fxi^\prime\rangle=K\langle\fxi\vert\fxi^\prime\rangle
{c_{N+1}+\sum^N_{\alpha=1}c_\alpha{\xi^\alpha}^*{\xi^\prime}^\alpha\over
1-\fxi^\dagger\fxi^\prime}\ .
\end{equation}
The trace formula is
\begin{eqnarray}
Z&=&e^{-ic_{N+1}KT}\lim_{M\to\infty}\int_{\rm
PBC}\prod^M_{i=1}d\mu\left(\fxi_i,
\fxi^\dagger_i\right)\exp\left[ K\left\{\sum^M_{k=1}\log\left(1-\fxi^\dagger_k
\fxi_k\right)\right.\right.\nonumber\\
&&\left.\left.-\sum^M_{k=1}\log\left(1-\sum^N_{\alpha=1}e^{-i\varepsilon
\mu_\alpha}{\xi^\alpha_k}^*\xi^\alpha_{k-1}\right)\right\}\right]\ ,
\label{ntrace}
\end{eqnarray}
where we have put $\mu_\alpha\equiv c_\alpha+c_{N+1}$.

The expansion parameter in this case is $1/K$.
The equations of motion
\begin{equation}
\cases{\displaystyle
-{\xi^\alpha_k\over1-\sum^N_{\beta=1}{\xi^\beta_k}^*\xi^\beta_k}
+{e^{-i\varepsilon\mu_\alpha}\xi^\alpha_{k-1}\over1-\sum^N_{\beta=1}
e^{-i\varepsilon\mu_\beta}{\xi^\beta_k}^*\xi^\beta_{k-1}}=0\cr\cr\cr
\displaystyle
-{{\xi^\alpha_k}^*\over1-\sum^N_{\beta=1}{\xi^\beta_k}^*\xi^\beta_k}
+{e^{-i\varepsilon\mu_\alpha}{\xi^\alpha_{k+1}}^*\over1-\sum^N_{\beta=1}
e^{-i\varepsilon\mu_\beta}{\xi^\beta_{k+1}}^*\xi^\beta_k}=0\ ,\cr
}
\end{equation}
have the solutions
\begin{equation}
\xi^\alpha_k={\xi^\alpha_k}^*=0\hspace{6ex}{\rm\ for\ all}\ k\
{\rm and\ for\ all}\ \alpha\ ,
\end{equation}
because the integration domain, $D_{N,1}$ does not contain $\infty$.

Put
\begin{equation}
\xi^\alpha_k\to\sqrt{\kappa}\xi^\alpha_k\ ,\ \left(\kappa\equiv{1\over K}
\right)\ ,
\end{equation}
to give
\begin{eqnarray}
Z&=&e^{-ic_{N+1}KT}\lim_{M\to\infty}\int\prod^M_{i=1}{\left[ d\fxi^\dagger_i
d\fxi_i\right]^N\over\pi^N}\nonumber\\
&&\times\exp\left[-\sum^M_{k=1}\left(\sum^N_{\alpha=1}
{\xi^\alpha_k}^*\xi^\alpha_k
-\sum^N_{\alpha=1}e^{-i\varepsilon\mu_\alpha}{\xi^\alpha}^*_k
\xi^\alpha_{k-1}\right)\right.\nonumber\\
&&+\sum^\infty_{n=1}\sum^M_{m=1}{\kappa^n\over n}
\left[-\sum^N_{\alpha=1}\alpha^n+\left( N+1\right)\left(
\sum^N_{\alpha=1}{\xi^\alpha_k}^*\xi^\alpha_k\right)^n\right.\nonumber\\
&&\left.\left.-{n\over
n+1}\left\{\left(\sum^N_{\alpha=1}{\xi^\alpha_k}^*\xi^\alpha_k
\right)^{n+1}-\left(\sum^N_{\alpha=1}e^{-i\varepsilon\mu_\alpha}
{\xi^\alpha_k}^*\xi^\alpha_k\right)^{n+1}\right\}\right]\right]\ .
\label{ntenkai}
\end{eqnarray}
Comparing (\ref{ntenkai}) with (\ref{tenkai}), we find the correspondence;
\begin{eqnarray}
K&\leftrightarrow&Q\ ,\nonumber\\
\kappa&\leftrightarrow&-\kappa\ .
\end{eqnarray}
Thus, without any explicit calculation, the WKB approximation is again
exact:
\begin{equation}
Z=e^{-ic_{N+1}KT}\prod^N_{\alpha=1}{1\over1-e^{-i\mu_\alpha T}}\ .
\label{nkekka}
\end{equation}

\section{Discussion}\label{sec:giron}

So far the WKB-exactness is shown in $CP^N$ and $D_{N,1}$.
In this section we try to comprehend its significance more seriously.
The Duistermaat-Heckman(D-H) theorem~\cite{DH} starts with that
let $M$ be a $2N$-dimensional symplectic manifold with the symplectic
form $\omega$
and a {\it torus} $T$ which acts on $M$ in a Hamiltonian way;
``Hamiltonian action'' designates that there is given a linear map
\begin{equation}
X\in{\bf t}\mapsto J_X\in C^\infty\left( M\right)\ ,\ {\bf t}\ :
\ {\rm Lie\ algebra\ of}\ T\ ,
\end{equation}
such that
\begin{enumerate}
\item for each $X\in{\bf t}$ the infinitesimal action of $X$ on $M$ is
generated by the Hamiltonian vector field $\tilde X$ of the function
$J_X$, which  satisfies
\begin{equation}
dJ_X=-i_{\tilde X}\omega\ .
\end{equation}
\item the functions $J_X(X\!\in\!{\bf t})$ are in involution.
\item there exists the momentum mapping of the Hamiltonian $T$-action.
$J\ :\ M\mapsto {\bf t}^*$, defined by
\begin{equation}
\langle X,J\left( m\right)\rangle=J_X\left( m\right),m\in M\ ,\ X\in{\bf t}\ .
\end{equation}
\end{enumerate}
It then says
\begin{flushleft}{\bf Theorem}:\end{flushleft}
\begin{equation}
\int_M{\omega^N\over N!}\exp\left(-\rho J_X\right)=
\sum_{m_c}{\exp\left\{-\rho J_X\left( m_c\right)\right\}\over
\rho^ND\left( m_c\right)}\ ,\ dH\left( m_c\right)=0\ ,\ m_c\in M\ ,
\end{equation}
where $D(m_c)$ is interpreted as the Gaussian determinant arising from the
saddle point approximation at the critical point $m_c$ and $\rho$ is
a real parameter.

The first example of the theorem is $(M,T)=({\bf C}^{N+1},T^{N+1})$, where
\begin{equation}
T^{N+1}\equiv\left\{ t=t\left(\theta_1,\cdots,\theta_{N+1}\right)=\left.
\pmatrix{e^{i\theta_1}&&\cr&\ddots&\cr&&e^{i\theta_{N+1}}\cr}\right\vert
\theta_1,\cdots,\theta_{N+1}\in{\bf R}\right\}\ .
\end{equation}
The symplectic structure of ${\bf C}^{N+1}$ is
\begin{eqnarray}
\omega&\equiv&id{\bf z}^\dagger\wedge d{\bf z}\ , \
{\bf z}\equiv\left( z^1,\cdots,z^{N+1}\right)^T\ ,\nonumber\\
\left\{ z^\alpha,{z^\beta}^*\right\}&=&-i\delta^{\alpha\beta}\ ,\
\left\{ z^\alpha,z^\beta\right\}=\left\{{z^\alpha}^*,{z^\beta}^*\right\}=0\ .
\end{eqnarray}
An action of $T^{N+1}$ on ${\bf C}^{N+1}$ is
\begin{equation}
\pmatrix{z^1\cr\vdots\cr z^{N+1}\cr}\in{\bf C}^{N+1}\mapsto
t\left(\theta_1,\cdots,\theta_{N+1}\right){\bf z}
=\pmatrix{e^{i\theta_1}z^1\cr\vdots\cr e^{i\theta_{N+1}}z^{N+1}\cr}\ ,
\end{equation}
and the Hamiltonian action reads
\begin{equation}
X=\pmatrix{i\theta_1&&\cr&\ddots&\cr&&i\theta_{N+1}\cr}\in{\bf t}\mapsto
J_X\left({\bf z}^\dagger,{\bf z}\right)=-i{\rm tr}\left({\bf z}{\bf z}^\dagger
X\right)=-i{\bf z}^\dagger X{\bf z}\in{\bf R}
\ .\label{shazou}
\end{equation}
The infinitesimal action of $T^{N+1}$ is generated by a vector field:
\begin{equation}
\tilde X=i\sum^{N+1}_{\alpha=1}\theta_\alpha\left( z^\alpha
{\partial\over\partial z^\alpha}-{z^\alpha}^*
{\partial\over\partial{z^\alpha}^*}\right)\ ,\
\left(\theta_\alpha\ne0\right)\ .
\end{equation}
For fixed $X$ a $U(1)$ action on ${\bf C}^{N+1}$,
\begin{equation}
\exp\left( a\tilde X\right) F\left({\bf z}^\dagger,{\bf z}\right)=
F\left({\bf z}^\dagger e^{-aX},e^{aX}{\bf z}\right)\ .
\end{equation}
The fixed point is found only at ${\bf z}=0$.
The following relation then holds:
\begin{equation}
\int_{{\bf C}^{N+1}}{\left[ d{\bf z}^\dagger d{\bf z}
\right]^{N+1}\!\!\!\!\!\!\!\!\over\pi^{N+1}}
\exp\left\{-\rho J_X\left({\bf z}^\dagger,{\bf z}\right)\right\}=
{1\over\rho^{N+1}\prod^{N+1}_{\alpha=1}\theta_\alpha}\ ,
\end{equation}
which is trivial because of the Gaussian integral.

The second example is $(M,T)=(CP^N,T^{N+1})$:
the symplectic structure is
\begin{eqnarray}
\omega\equiv i{\rm tr}\left( P^{\left(N+1\right)}dP^{\left(N+1\right)}\wedge
dP^{\left(N+1\right)}\right)=
{i\over1+\fxi^\dagger\fxi}d\fxi^\dagger\wedge\left({\bf 1}_N
+\fxi\fxi^\dagger\right)^{-1}d\fxi\ ,\nonumber\\
P^{\left( N+1\right)}\equiv{1\over1+\fxi^\dagger\fxi}
\pmatrix{\fxi\fxi^\dagger&\fxi^\dagger\cr\fxi&1\cr}\ ,\
\fxi\equiv\left( \xi^1,\cdots,\xi^N\right)^T\in{\bf C}^N\ ,\nonumber\\
\left\{\xi^\alpha,{\xi^\beta}^*\right\}=-i\left(1+\fxi^\dagger\fxi\right)
\left(\delta^{\alpha\beta}+{\xi^\alpha}^*\xi^\beta\right)\
,\ \left\{\xi^\alpha,\xi^\beta\right\}=\left\{{\xi^\alpha}^*,
{\xi^\beta}^*\right\}=0\ .
\end{eqnarray}
A $T^{N+1}$-action reads
\begin{equation}
\pmatrix{\xi^1\cr\vdots\cr \xi^N\cr}\in{\bf C}^N\mapsto
t\left(\theta_1,\cdots,\theta_{N+1}\right)\fxi=
\pmatrix{e^{i\left(\theta_1-\theta_{N+1}\right)}\xi^1\cr\vdots\cr
e^{i\left(\theta_N-\theta_{N+1}\right)}\xi^N\cr}\ ,
\end{equation}
and the Hamiltonian action is
\begin{eqnarray}
X=\pmatrix{i\theta_1&&\cr&\ddots&\cr&&i\theta_{N+1}\cr}\in{\bf t}&\mapsto&
J_X\left(\fxi^\dagger,\fxi\right)=-i{\rm tr}\left( P^{\left( N+1\right)}
X\right)\nonumber\\
&&={\theta_{N+1}+\sum^N_{\alpha=1}\theta_\alpha{\xi^\alpha}^*\xi^\alpha
\over1+\fxi^\dagger\fxi}\in{\bf R}\ .
\end{eqnarray}
A vector field, for fixed $X$,
\begin{equation}
\tilde X=i\sum^N_{\alpha=1}\left(\theta_\alpha-\theta_{N+1}\right)
\left(\xi^\alpha{\partial\over\partial\xi^\alpha}-{\xi^\alpha}^*
{\partial\over\partial{\xi^\alpha}^*}\right)\ ,\
\left(\theta_\alpha\ne\theta_\beta\ {\rm for}\ \alpha\ne\beta\right)\ ,
\end{equation}
generates a $U(1)$ action on $CP^N$,
\begin{eqnarray}
&&\exp\left( a\tilde X\right) F\left(\fxi^\dagger,\fxi\right)=
F\left(\fxi^\dagger e^{-a\hat X},e^{a\hat X}\fxi\right)\ ,\nonumber\\
&&\hat X\equiv\pmatrix{i\left(\theta_1-\theta_{N+1}\right)&&\cr&\ddots&\cr
&&i\left(\theta_N-\theta_{N+1}\right)\cr}\ .
\end{eqnarray}
The fixed point is found at $\fxi=0$.
There are $N$ other parameterizations in $CP^N$ such that
\begin{eqnarray}
P^{\left(\alpha\right)}&\equiv&U^{\left(\alpha,N+1\right)}P^{\left(
N+1\right)}
{U^{\left(\alpha,N+1\right)}}^\dagger\ ,
\ U^{\left(\alpha,N+1\right)}\in U\left( N+1\right)\ ,\nonumber\\
\left( U^{\left(\alpha,N+1\right)}\right)_{\mu\nu}&\equiv&\delta_{\mu\nu}
-\delta_{\mu\alpha}\delta_{\nu\alpha}-\delta_{\mu,N+1}\delta_{\nu,N+1}
+\delta_{\mu,N+1}\delta_{\nu\alpha}+\delta_{\mu\alpha}\delta_{\nu,N+1}
\ ,\nonumber\\
&&\alpha=1,\cdots,N\ ,
\end{eqnarray}
and the fixed point, $\fxi=0$, is common in every case.
Therefore
\begin{equation}
\int_{CP^N}{\left[ d\fxi^\dagger d\fxi\right]^N\over\pi^N
\left(1+\fxi^\dagger\fxi\right)^{N+1}}
\exp\left\{-\rho J_X\left(\fxi^\dagger,\fxi\right)\right\}=
\sum^{N+1}_{\alpha=1}{e^{-\rho\theta_\alpha}\over\rho^N\prod_{\beta\ne\alpha}
\left(\theta_\beta-\theta_\alpha\right)}\ .\label{kotencpn}
\end{equation}

The third example is a case that does not fit the D-H requirement:
$(M,T)=(T^2,T^2)$. The symplectic structure is
\begin{eqnarray}
&&\omega\equiv d\psi\wedge d\varphi\ ,\
p\equiv\left( e^{i\psi},e^{i\varphi}\right)\in S^1\times S^1\ ,\nonumber\\
&&\left\{\psi,\varphi\right\}=-1\ ,
\end{eqnarray}
a $T^2$-action is
\begin{equation}
\left( e^{i\psi},e^{i\varphi}\right)\in S^1\times S^1\mapsto t\left(\theta_1,
\theta_2\right) p=\left( e^{i\left(\psi+\theta_1\right)},
e^{i\left(\varphi+\theta_2\right)}\right)\ ,
\end{equation}
and the Hamiltonian action\footnote{$J_X$ can be regarded as Hamiltonian
locally, but not globally\cite{AM}.} is
\begin{equation}
X=\pmatrix{i\theta_1&0\cr0&i\theta_2\cr}\in{\bf t}\mapsto J_X\left( p\right)
=\theta_2\psi-\theta_1\varphi\ .
\end{equation}
A vector field, for fixed $X$,
\begin{equation}
\tilde X=\theta_1{\partial\over\partial\psi}+\theta_2
{\partial\over\partial\varphi}\ ,
\end{equation}
generates a $U(1)$ action on $T^2$ such that
\begin{equation}
\exp\left( ia\tilde X\right) F\left( p\right)=F\left( t\left( a\theta_1,
a\theta_2\right) p\right)\ .\label{vectorsayou}
\end{equation}
The facts that there is no fixed point in (\ref{vectorsayou}) and that the
integral is given by
\begin{equation}
\int_{T^2}d\psi d\varphi\exp\left\{-\rho J_X\left( p\right)\right\}=
{1\over\rho^2\theta_1\theta_2}\left( e^{2\pi\rho\theta_1}-1\right)
\left(1-e^{-2\pi\rho\theta_2}\right)\ ,
\end{equation}
clearly show that the D-H formula breaks in this case.
The reason is that $\{ J_X,J_{X^\prime}\}\ne0$, where
$J_X(p)=\theta_2\psi-\theta_1\varphi$ and  $J_{X^\prime}(p)=
\theta^\prime_2\psi-\theta^\prime_1\varphi$,
does not meet the assumption 2.

The situations can be viewed from a different stand: the first example
is nothing but a Gaussian integral and the second Hamiltonian is a
perfect Morse function on $CP^N$ (Even more it is a perfect Morse
function\cite{FK}).
Both Hamiltonians are invariant under $U(1)$ transformation,
$z\mapsto e^{ia}z$, $\xi\mapsto e^{ia}\xi$, which is closely
related to the assumption 2.
The third one is neither $U(1)$ invariant nor a Morse function.
Even if we adopt a Morse function as the Hamiltonian, the result
does not fulfill the D-H theorem\cite{KNST}:
\begin{equation}
\int_{T^2}{d\psi d\varphi\over\left(2\pi\right)^2}\exp\left\{-\rho\left(
a\cos\psi+b\cos\varphi\right)\right\}
=I_0\left(\rho a\right) I_0\left(\rho b\right)\ ,\label{ttmorse}
\end{equation}
where $I_0(z)$ is the modified Bessel function.
The difference between first two and the third example can be discerned
much clearly by introducing new variables, $z_1=\vert z_1\vert e^{i\psi}$,
$z_2=\vert z_2\vert e^{i\varphi}$,
to rewrite (\ref{ttmorse}) such that
\begin{eqnarray}
{\rm L.H.S.\ of\ }(\ref{ttmorse})&=&\int_{{\bf C}^2}{d{z_1}^*dz_1\over\pi}
{d{z_2}^*dz_2\over\pi}
\delta\left(\vert z_1\vert^2-1\right)\delta\left(\vert z_2\vert^2-1\right)
\nonumber\\
&&\times\exp\left[-\rho\left\{{a\over2}\left( z_1+{z_1}^*\right)+{b\over2}
\left( z_2+{z_2}^*\right)\right\}\right]\nonumber\\
&=&\int^\infty_{-\infty}{d\lambda_1d\lambda_2\over\left(2\pi\right)^2}
\int_{{\bf C}^2}{d{z_1}^*dz_1\over\pi}{d{z_2}^*dz_2\over\pi}\nonumber\\
&&\times\exp\Bigg[-\rho\left\{{a\over2}\left( z_1+{z_1}^*\right)+{b\over2}
\left( z_2+{z_2}^*\right)\right\}\nonumber\\
&&+i\lambda_1\left(\vert z_1\vert^2-1\right)
+i\lambda_2\left(\vert z_2\vert^2-1\right)\Bigg]\ .\label{ttkosoku}
\end{eqnarray}
Further the right hand side of (\ref{kotencpn}) can be expressed as
\begin{equation}
\left(\ref{kotencpn}\right)=\int^\infty_{-\infty}{d\lambda\over2\pi}
\int_{{\bf C}^{N+1}}{\left[ d{\bf z}^\dagger d{\bf z}
\right]^{N+1}\!\!\!\!\!\!\!\!\over\pi^{N+1}}
\exp\left[-\rho{\bf z}^\dagger X{\bf z}+i\lambda
\left({\bf z}^\dagger{\bf z}-1\right)\right]\ ,\label{kotencpnkosoku}
\end{equation}
under these expressions we see that $\lambda$-integral brings trivial
(flat) manifolds to nontrivial ones through the condition ${\bf z}^\dagger
{\bf z}=1$ which thus can be designated as the constraints.

Now the differences are
\begin{itemize}
\item the Hamiltonian of (\ref{kotencpnkosoku}) is bilinear but that of
(\ref{ttkosoku}) is not, and
\item in (\ref{kotencpnkosoku}) the Poisson bracket between the Hamiltonian
and the constraint vanishes while in (\ref{ttkosoku}) they do not.
\end{itemize}

In (\ref{kotencpnkosoku}) the Gaussian integral with respect to $
{\bf z}^\dagger{\bf z}$
can be performed to be
\begin{equation}
\left(\ref{kotencpnkosoku}\right)=\int^\infty_{-\infty}d\lambda e^{-i\lambda}
\prod^{N+1}_{\alpha=1}{1\over\rho\theta_\alpha-i\lambda}\ .\label{gaussato}
\end{equation}
The role of the $\lambda$-integral is to sum up the contributions of residues
to give
\begin{equation}
\left(\ref{gaussato}\right)=\sum^{N+1}_{\alpha=1}e^{-\rho\theta_\alpha}
\prod_{\beta\ne\alpha}{1\over\rho\left(\theta_\beta-\theta_\alpha\right)}\ ,
\end{equation}
which coincides with the result of the D-H theorem.
The conditions of the saddle point in (\ref{kotencpnkosoku}) are
\begin{equation}
\cases{\left(\rho X-i\lambda\right){\bf z}=0\ ,\cr
{\bf z}^\dagger\left(\rho X-i\lambda\right)=0\ ,\cr
{\bf z}^\dagger{\bf z}-1=0\ ,}\label{saddle}
\end{equation}
whose solutions ${\bf z}^c$ are the eigenvectors of $\rho X$ with the
eigenvalues
$i\lambda=\rho\theta_\alpha$'s(See (\ref{shazou})).
Thus we find the poles in (\ref{gaussato}) originate from the critical points
of the system.

Now apply this point of view to the case in the preceding sections, to wit,
to the quantum version of the D-H theorem.
For the sake of simplicity we concentrate on the compact case.
The trace formula (\ref{teigitr}) can be given in terms of the oscillators
(\ref{ztaninobunkai}) as
\begin{equation}
Z=\int{\left[ d{\bf z}^\dagger d{\bf z}\right]^{N+1}\!\!\!\!\!\!\!\!
\over\pi^{N+1}}
\langle{\bf z}\vert P_Qe^{-iHT}\vert{\bf z}\rangle\ ,\label{zdekaku}
\end{equation}
which becomes
\begin{eqnarray}
Z&=&\lim_{\delta\to0}\int^{2\pi}_0{d\lambda\over2\pi}e^{-i\lambda Q}
\int{\left[ d{\bf z}^\dagger d{\bf z}\right]^{N+1}\!\!\!\!\!\!\!\!
\over\pi^{N+1}}
\langle{\bf z}\vert\exp\left[-i\sum^{N+1}_{\alpha=1}a^\dagger_\alpha
\left\{ c_\alpha T-\left(\lambda+i\delta\right)\right\} a_\alpha\right]
\vert{\bf z}\rangle\nonumber\\
&=&\lim_{\delta\to0}\int^{2\pi}_0{d\lambda\over2\pi}e^{-i\lambda Q}
\lim_{M\to\infty}
\int{\left[ d{\bf z}^\dagger d{\bf z}\right]^{N+1}\!\!\!\!\!\!\!\!
\over\pi^{N+1}}
\langle{\bf z}\vert\left[1-i{1\over M}\sum^{N+1}_{\alpha=1}a^\dagger_\alpha
\left\{ c_\alpha T-\left(\lambda+i\delta\right)\right\} a_\alpha\right]^M\!\!
\vert{\bf z}\rangle\nonumber\\
&=&\lim_{\delta\to0}\int^{2\pi}_0{d\lambda\over2\pi}
e^{-i\lambda Q}\lim_{M\to\infty}\int_{\rm PBC}\prod^M_{i=1}
{\left[ d{\bf z}_i^\dagger d{\bf
z}_i\right]^{N+1}\!\!\!\!\!\!\!\!\over\pi^{N+1}}
\nonumber\\
&&\times\exp\left[-\sum^M_{k=1}
\sum^{N+1}_{\alpha=1}\left\{{z^\alpha_k}^*z^\alpha_k-e^{-i\varepsilon
\left( c_\alpha-{\lambda+i\delta\over T}\right)}
{z^\alpha_k}^*z^\alpha_{k-1}\right\}\right]\ .\label{trofharm}
\end{eqnarray}
Here $\varepsilon$ is defined by (\ref{risanka}) and the definition
of the exponential function has been used from the first to the
second line then the resolutions of unity (\ref{ztaninobunkai})
has been inserted into the third line to convert $a_\alpha^\dagger$,
$a_\alpha$ to $c$-numbers ${z^\alpha}^*$, $z^\alpha$.
The important relation we have employed is
\begin{equation}
\left[ H,P_Q\right]=0\ .
\end{equation}
In (\ref{trofharm}) the integrals with respect to ${\bf z}^\dagger$,
 ${\bf z}$ is Gaussian, as in (\ref{kotencpnkosoku}), to give
\begin{eqnarray}
Z&=&\lim_{\delta\to0}\int^{2\pi}_0{d\lambda\over2\pi}e^{-i\lambda Q}
\prod^{N+1}_{\alpha=1}{1\over1-e^{-ic_\alpha T+i\left(\lambda+i\delta\right)}}
\nonumber\\
&=&\lim_{\delta\to0}\oint_{\left\vert w\right\vert=1}{dw\over2\pi}w^{Q+N}
\prod^{N+1}_{\alpha=1}{1\over w-e^{-ic_\alpha T-\delta}}\nonumber\\
&=&\sum^{N+1}_{\alpha=1}e^{-iQc_\alpha T}\prod_{\beta\ne\alpha}
{1\over1-e^{-i\left( c_\beta-c_\alpha\right) T}}\ ,\label{zkekka}
\end{eqnarray}
where the $\lambda$-integral has been transformed to the contour integral in
the second line.
Note that the $w$-integral picks up the $N+1$ poles to give the exact result.

Here we can recognize the role of the poles in (\ref{zkekka}) as the critical
points as above: the equations of motion reads
\begin{equation}
\cases{\displaystyle z^\alpha_k-e^{-i\varepsilon\left( c_\alpha-{\lambda\over
T}
\right)}z^\alpha_{k-1}=0\ ,\cr
\displaystyle{z^\alpha_k}^*-e^{i\varepsilon\left( c_\alpha-
{\lambda\over T}\right)}{z^\alpha_{k-1}}^*=0\ ,\cr
\displaystyle {1\over M}\sum^M_{k=1}\sum^{N+1}_{\alpha=1}
e^{-i\varepsilon\left( c_\alpha-{\lambda\over T}\right)}
{z^\alpha_k}^*z^\alpha_{k-1}-Q=0\ . }\label{kaiz}
\end{equation}
Therefore the solutions
\begin{equation}
z^\alpha_k=e^{-ik\varepsilon\left( c_\alpha-{\lambda\over T}\right)}
z^\alpha_0\ ,
\end{equation}
combined with the periodic boundary condition,
\begin{equation}
z^\alpha_M=z^\alpha_0\ ,
\end{equation}
gives
\begin{equation}
\left\{1-e^{-i\left( c_\alpha T-\lambda\right)}\right\} z^\alpha_0=0\ ,
\end{equation}
or in the vector notation
\begin{equation}
\left\{1-e^{-i\left( H T-\lambda\right)}\right\}{\bf z}_0=0\ ,
\end{equation}
where $H$ has been given by (\ref{hamiltonian}).
${\bf z}_0\!=\!0$ does not meet the third relation (constraint)
in (\ref{kaiz}) thus
the only remaining case is that ${\bf z}_0$ is the eigenvector of
$e^{-iHT}$ with the eigenvalue $e^{-i\lambda}$.
There holds a complete analogy between the above ``classical''
(\ref{kotencpnkosoku}) and its quantum version (\ref{trofharm}):
the pole structure in the contour integral (\ref{gaussato}) or
(\ref{zkekka}) corresponds to the eigenvalue equation (\ref{saddle})
or (\ref{kaiz}).
The former picks up eigenvalues of {\sl Hermitian operator} while the
latter does those of {\sl unitary operator}.

Now we summarize our observation:
classical system met with the D-H theorem could be generalized easily
to a corresponding quantum counterpart if we regard the target manifold
as the constraint system embedded in a (trivial) manifold.
The situation would be if
\begin{enumerate}
\item Hamiltonian $H$ is bilinear of creation and annihilation
operators, and
\item constraint $P$ is commutable with Hamiltonian,
$[H,P]=0$,
\end{enumerate}
then the system is WKB-exact.
The first condition is necessary; since there is a case which
has no higher order corrections but might not match the exact
result: for example $H=({\bf a}^\dagger{\bf a})^2,P=P({\bf a}^\dagger
{\bf a})\ {\rm (Polynomial\ in\ {\bf a}^\dagger{\bf a})}$.

The generalization to more generic cases such as Grassmannian is now
under investigation.

\end{document}